\documentstyle[aps,prl]{revtex}
\begin{document}
\draft
\title{Incommensurate Charge and Spin Fluctuations in d-wave Superconductors}
\author{Hae-Young Kee$^{1}$ and Yong Baek Kim$^{2}$}
\address{
${}^1$ Department of Physics, Rutgers University, Piscataway, NJ
08855-0849\\
${}^2$ Department of Physics, The Pennsylvania State University,
University Park, PA 16802}

\date{\today}
\maketitle

\begin{abstract}
We show analytic results for the irreducible charge and spin 
susceptibilities,
$\chi_0 (\omega, {\bf Q})$, where ${\bf Q}$ is the momentum 
transfer between the nodes in d-wave superconductors.
Using the BCS theory and a circular Fermi surface, we find that
the singular behavior of the irreducible charge susceptibility leads to
the dynamic incommensurate charge collective modes.
The peaks in the charge structure factor occur at a set of wave
vectors which form an ellipse around ${\bf Q}_{\pi}=(\pi,\pi)$ 
and ${\bf Q}_0=(0,0)$
in momentum space with momentum dependent spectral weight.
It is also found that, due to the  non-singular irreducible 
spin susceptibility,  an extremely strong interaction via random phase
approximation is required to support 
the magnetic peaks near ${\bf Q}_{\pi}$. Under certain conditions,
the peaks in the magnetic structure factor occur near
${\bf Q}=(\pi,\pi (1 \pm \delta))$ and $(\pi (1 \pm \delta),\pi)$. 
\end{abstract}


The neutron scattering measurements provide the direct information
about the wavevector and frequency dependence of the dynamic
spin susceptibility.
On the other hand  the inelastic x-ray or electron scattering
can measure the dynamic charge susceptibility.
These informations are particularly important
in the cuprate superconductors because of the
intimate interplay between the spin and charge dynamics
which may be related to the mechanism of the high temperature
superconductivity.

The neutron scattering experiments on $La_{2-x}Sr_xCuO_4(LSCO)$\cite{cheong,mook}
found the 
inelastic incommensurate peaks near ${\bf Q}_{\pi}=(\pi,\pi)$ in 
the superconducting state as well as in the normal 
state\cite{cheong}.
The collective modes occur at 
${\bf Q}=(\pi,\pi (1 \pm \delta))$ and
$(\pi (1 \pm \delta),\pi)$ 
for low frequencies.
Recently, a lot of effort was put forward to investigate the
spin dynamics in $YBa_2Cu_3O_{7-y}(YBCO)$\cite{dai,keimer}.
In earlier works, the incommensurate peaks were reported
in the scan along the zone diagonal direction\cite{dai}.
Later it was found that the locations of the higher intensity
peaks are the same as those of $LSCO$; $(\pi,\pi(1 \pm \delta))$
and $(\pi(1 \pm \delta),\pi)$\cite{dai2}.

On the theoretical front,
there have been two different approaches to explain the incommensurate
peaks.
The numerical studies of  one- and three-band Hubbard models found that
the incommensurate peaks occur as the dynamical response of a spatially
homogeneous system  with a nearly nested Fermi surface.\cite{bulut,qimiao}
It was also pointed out that the nesting peaks are at $(\pi(1 \pm
\delta),\pi(1\pm \delta))$ and
$(\pi(1\pm \delta), \pi(1\mp \delta))$ in $YBCO$
due to the $45$ degree rotation of the Fermi surface.
\cite{qimiao}
From a different viewpoint, it was proposed that the magnetic 
incommensurate peaks
are induced by the dynamic charge stripes in a spatially 
inhomogeneous system.\cite{kivelson}
The static charge ordering can occur in the relatively bad metal 
due to the pinning of the charge stripes.
On the other hand, the dynamic fluctuations of the charge
stripes result in the phase coherence of the superconducting state.

Recently, Tranquada {\it et al} found, in the neutron diffraction
measurements, the incommensurate elastic magnetic peaks at 
${\bf Q}=(\pi,\pi (1 \pm \delta))$ and
$(\pi (1 \pm \delta),\pi)$ in $La_{1.6-x}Nd_{0.4} Sr_x CuO_4$.\cite{tranquada}
They interpreted these peaks as the evidence for the static stripe order 
of charge and spins.
It was also shown that the superconducting transition temperature 
increases as the peak splitting parameter, $\delta$, increases. 
This may imply that the charge and spin stripes are intimately related to
the superconductivity.

These elastic and inelastic neutron scattering experments motivated us 
to investigate
the incommensurate charge and spin fluctuations 
in the superconducting state.
In particular, since the inelastic experiment on $YBCO$ found the 
incommensurate peaks at the same positions as those of $LSCO$, 
these may not be related to the details of the Fermi surface shape.
Thus, as a first step, it is worthwhile to understand the charge 
and spin dynamics within the BCS theory with a simple Fermi surface.
This would provide an important information which should be compared with any 
new theoretical proposal-either Fermi surface effect or the charge stripes.

In this paper, we compute analytically the irreducible charge and spin 
susceptibilities in the BCS d-wave superconducting state with a circular
Fermi surface.
We found that 
the incommensurate charge collective modes
occur at a set of wave vectors.
The wave vectors form an ellipse around ${\bf Q}_{\pi}=(\pi,\pi)$ 
in momentum space.
The spectral weight also depends on the wave vectors.
It is found that the incommensurate spin collective mode can occur
near ${\bf Q}=(\pi,\pi (1 \pm \delta))$ and
$(\pi (1 \pm \delta),\pi)$ only when an extremely strong interaction
is assumed via the random phase approximation.
Using the analytic form of the irreducible spin susceptibility,
 we also examine the NMR relaxation
rate $1/T_1$ for low temperatures and
found that $(T_1 T)^{-1} \sim T^2/(\Delta E_F)^2$.
We also analyze the quasiparticle lifetime for small frequencies
at the nodes; 
$1/\tau \sim V^2\omega^3/(\Delta E_F)^2$, where $V$ is the
interaction.

Let us consider the simplest  model for the electronic energy
with a circular Fermi-surface of radius $k_F$.
\begin{equation}
\xi_{\bf k}=\epsilon_{\bf k}-\mu =  \frac{k_x^2}{2m}
+ \frac{k_y^2}{2m}-\frac{k_F^2}{2 m} \ , 
\label{model}
\end{equation}
The lowest order charge, $\chi_0^c$, and spin susceptibilities, $\chi_0^s$, 
for  momentum {\bf Q} and energy $\omega$ at $T=0$ are given by 
\begin{equation}
\chi_0^{c}(\omega,{\bf Q})=
-\frac{1}{2} \sum_{\bf k}
 \left( 1-\frac{\xi_{{\bf k}+{\bf Q}}\xi_{\bf k}
-\Delta_{{\bf k}+{\bf Q}} \Delta_{\bf k}}
{E_{{\bf k}+{\bf Q}}E_{\bf k}} \right)
\left( \frac{1}{\omega+E_{{\bf k}+{\bf Q}}+E_{\bf k}+i\eta}
- \frac{1}{\omega-E_{{\bf k}+{\bf Q}}-E_{\bf k}+i\eta} \right) \ ,
\label{charge}
\end{equation}
 
\begin{equation}
\chi_0^{s}(\omega,{\bf Q})=
\frac{1}{2} \sum_{\bf k}
 \left( 1-\frac{\xi_{{\bf k}+{\bf Q}}\xi_{\bf k}
+\Delta_{{\bf k}+{\bf Q}} \Delta_{\bf k}}
{E_{{\bf k}+{\bf Q}}E_{\bf k}} \right)
\left( \frac{1}{\omega+E_{{\bf k}+{\bf Q}}+E_{\bf k}+i\eta}
- \frac{1}{\omega-E_{{\bf k}+{\bf Q}}-E_{\bf k}+i\eta} \right) \ ,
\label{spin}
\end{equation}
where $E_{\bf k}=\sqrt{\xi_{\bf k}^2+\Delta_{\bf k}^2}$
with $\Delta_{\bf k}=\Delta \cos{2 \phi}$.
The different coherence factors in the charge and spin 
susceptibilities 
in Eq. (\ref{spin}) come from the fact that 
the  magnetic scattering is  odd 
with respect to the time reversal symmetry while the charge scattering
is even.

Let us exmaine the momentum transfer near
${\bf Q}_{\pi}=(\sqrt{2} k_F,\sqrt{2} k_F)$ or $(\pi,\pi)$\cite{note1}
shown in Fig. 1.
Near the nodes, we  expand the electronic dispersion relation and
the amplitude of the gap.
We find the following for $|{\bf p}|+|{\bf q}| < k_F/\sqrt{2}$.
\begin{eqnarray}
\xi_{\bf k} &\simeq & -\sqrt{2} v_F (p_x+p_y)  \ , \hspace{2cm}
\Delta_{\bf k}  \simeq  -\sqrt{2} \Delta (p_x-p_y)/k_F \ ,\nonumber\\
\xi_{{\bf k}+{\bf Q}} &\simeq & \sqrt{2} v_F (p_x+q_x+p_y+q_y) \ ,  \  
\Delta_{{\bf k}+{\bf Q}}  \simeq  \sqrt{2} \Delta
(p_x+q_x-p_y-q_y)/k_F \ .
\end{eqnarray}
Here ${\bf p}={\bf k}-{\bf k}_0$ and ${\bf q}={\bf Q}-{\bf Q}_{\pi}$,
where ${\bf k}_0$ denotes the node of d-wave superconductors.
Note also that one obtains the same dispersion relations near $(\pi,\pi)$
in the tight binding model with $ 1/m=ta^2$ and $\sqrt{2}
k_F=\pi/2a$\cite{note1,wen}, where $t$ is the hopping amplitude and 
$a$ is the lattice spacing.
Then, the Dirac dispersion for the quasiparticle is found near the nodes;
\begin{eqnarray}
E_{\bf k}& = & \sqrt{2 E_F^2 p_+^2+2\Delta^2 p_-^2} \ ,\nonumber\\
E_{{\bf k}+{\bf q}} &=& \sqrt{2 E_F^2 (p_+ + q_+)^2+ 2 
\Delta (p_- +q_-)^2} \ ,
\label{dirac}
\end{eqnarray}
where 
\begin{equation}
p_{\pm}  =  (p_x \pm p_y)/k_F \ , \ \ \ \  
q_{\pm} = (q_x \pm q_y)/k_F \ .  
\end{equation}

Evaluating Eq.\ref{charge} with Eq.\ref{dirac}, 
the imaginary part of the charge susceptibilities
is found for $\omega_q \le \omega < \Delta$,
\begin{equation}
{\rm Im} \chi^c_0 (\omega, {\bf Q}) =\frac{1}{32 \Delta E_F}
 \frac{\omega^2- (\sqrt{2} E_F q_+)^2}
{\sqrt{\omega^2-\omega_q^2}} \ ,
\label{ellipse}
\end{equation}
where 
\begin{equation}
\omega_q^2=2 E_F^2 q_+^2+ 2\Delta^2 q_-^2 \ .
\end{equation}
It can be seen in Eq. (\ref{ellipse}) that  the charge 
susceptibility diverges as $\omega$ approaches $\omega_q$ and 
has the following scaling form;
\begin{equation}
{\rm Im} \chi^c_0 (\omega, {\bf Q}) = 
\frac{\omega}{32 \Delta E_F} f \left( \frac{\sqrt{2}E_F q_+}{\omega},
\frac{\sqrt{2} \Delta q_-}{\omega} \right) \ ,
\end{equation}
where
\begin{equation}
f(a,b)= \frac{1-a^2}{\sqrt{1-a^2-b^2}} \ .
\label{scale}
\end{equation}
Note that the incommensurate collective modes
occur with a set of wave vectors.
The wave vectors form an ellipse around ${\bf Q}_{\pi}=(\pi,\pi)$
in momentum space for the energy $\omega_q$.
The essentricity of the ellipse is determined by the ratio of
$\Delta$ and $E_F$.
Since $E_F$ is typically larger than $\Delta$ ($\frac{E_F}{\Delta}=5 \sim 10$),
the ellipse is elongated along the direction perpendicular to 
the zone diagonal. 
This result is a consequence of the
anisotropic Dirac dispersion relation of the quasiparticles 
near the node.
The Eq. (\ref{scale}) implies that the spectral weight of the
incommensurate collective mode depends on the wave vectors.
In particular, the spectral weight of the collective mode
vanishes at two points in the zone diagonal direction.
It is found that the charge susceptibility in the random 
phase approximation, 
$\chi^c(\omega, {\bf Q}) = 
\chi^c_0(\omega, {\bf Q}) / [1 + V(\omega, {\bf Q}) 
\chi^c_0(\omega, {\bf Q})]$,
is almost the same as the irreducible susceptibility, 
$\chi^c_0(\omega, {\bf Q})$,
for quite large range of the interaction strength 
if one assumes  $V(\omega, {\bf Q})=V $.
If  $E_F \sim 10 \Delta$, $V \sim  E_F$, and 
$\Delta \sim 30 meV$ , the correction to the position of the 
singularity, $\omega_q \sim 20 meV$, is order of $10^{-2} meV$.
Thus $\chi^c_0$ is enough to describe the charge susceptibility
as far as the most singular part is concerned.
  
In the lattice, the Umklapp scattering is present and it leads
to an additional contribution to the susceptibilty.
This contribution provides another collective modes
with the energy,  ${\tilde \omega}_q =2 E_F^2 q_-^2+2 \Delta q_+^2$.
As a result, the imaginary part of the susceptibility becomes
\begin{equation}
{\rm Im} \chi^c_{u} (\omega, {\bf Q}) =\frac{1}{32 \Delta E_F}
\left [  \frac{\omega^2-(\sqrt{2} E_F q_+)^2}
{\sqrt{\omega^2-\omega_q^2}} +
 \frac{\omega^2-(\sqrt{2} E_F q_-)^2}
{\sqrt{\omega^2-{\tilde \omega}_q^2}} \right ] \ .
\label{umklapp}
\end{equation}
One can see that the shape of the structure factor
depends on the ratio of the Fermi energy, $E_F$, and
the maximum amplitude of the gap, $\Delta$.
In Fig. 2, we show the imaginary  part of the charge susceptibility
for $E_F/\Delta=7$ and $2$.
It is usually assumed that the ratio of the Fermi energy
and the gap is order of $5 \sim 10$ because of the short
coherence length though it has  not been confirmed.
In a recent tunneling experiment, it was claimed 
that the ratio might be order of one in the case of 
$Bi_2Sr_2CaCu_2O_{8+y}(BiSCCO)$\cite{fischer}.
As  $E_F/\Delta$ gets bigger, the anisotropy of the structure factor
becomes stronger as shown in Fig. 2.
If one could detect the charge collective modes by inelastic x-ray
or electron scattering, it would provide the ratio of the Fermi energy
and the gap of the cuprates.

The charge susceptibility in the real space is  found from the Fourier
transform of Eq. (\ref{ellipse}) as
\begin{equation}
{\rm Im} \chi^c_0(\omega, x,y)= \frac{1}{64 \sqrt{2\pi} \Delta E_F}
\frac{(\omega  r)^{3/2}}{ r^5} [ (\omega r) r_-^2
J_{\frac{1}{2}} (\omega r)+(2r_+^2-r_-^2) J_{\frac{3}{2}} (\omega  r)],
\end{equation}
where $J_{\nu}(x)$ is the Bessel function and $r^2=r_+^2+r_-^2$. Here
$r_+=(x+y)/(2 \sqrt{2} E_F)$ and $r_-=(x-y)/(2 \sqrt{2} \Delta)$.
Since the periodicity of the  oscillation depends on the ratio
$E_F/\Delta$, the shape of the structure factor in the real space
is enlongated along the $r_-$ direction for $E_F > \Delta$ 
shown in Fig. 3.

Let us now study 
the spin susceptibility for the momentum transfer 
near ${\bf Q}_{\pi} = (\pi,\pi)$.
We  found 
\begin{equation}
{\rm Im} \chi^s_0 (\omega, {\bf Q}) =\frac{1}{16 \Delta E_F}
\sqrt{\omega^2-\omega_q^2} \ ,
\label{ellipse2}
\end{equation}
It can be clearly seen in Eqs. (\ref{ellipse}) and (\ref{ellipse2}) 
that the spin susceptibility is smooth while the charge susceptibility
diverges as mentioned above.
The difference between the charge and  spin susceptibilities
comes from the coherence factors.
The spin susceptibility has the maximum at ${\bf Q}_{\pi}$ ($\omega_q=0$)
for a given frequency $\omega$ and decreases 
as $\omega_q$ approaches $\omega$.
One can show that when  $\omega_q$ becomes larger than $\omega$ the
imaginary part of the irreducible spin susceptibility is negligible.
On the other hand, the real part of the irreducible spin susceptibility
is $(16\Delta E_F)^{-1} \sqrt{\omega_q^2-\omega^2}$ for $\omega_q > \omega$.

As a result, the spin susceptibility in the random 
phase approximation, 
$\chi^c(\omega, {\bf Q}) = 
\chi^c_0(\omega, {\bf Q}) / [1 + J(\omega, {\bf Q}) 
\chi^c_0(\omega, {\bf Q})]$,
can support the collective modes if 
the interaction $J(\omega, {\bf Q}) = J > 16 E_F$.
The shape of the spin structure factor depends on the ratio 
$E_F/\Delta$.
As mentioned above, the Umklapp scattering provides the additional
contribution to the spin susceptibility.
After including the Umklapp scattering, we find the following results. 
The positions of the magnetic peaks are  near
$(\pi,\pi(1 \pm \delta))$ and $(\pi(1 \pm \delta),\pi)$,
where $\delta$ (in units of $2\pi/a$) is given by
\begin{equation}
\delta = \frac{1}{2} \frac {\sqrt{\omega^2+(16\Delta E_F/J)^2}}
{\sqrt{E_F^2 + \Delta^2}} \ .
\label{posdel}
\end{equation}
If $\omega < (16 \Delta E_F/J)$ and $E_F/\Delta \sim 5-10$, 
the peak splitting parameter,
$\delta$, is order of $0.1$ and alomost independent of the frequency. 
In the experiments, 
$\delta$ seems to be frequency independent\cite{cheong,mook}. 
Our results imply that 
the interaction strength, $J$, and the ratio of the Fermi energy and
the amplitude of the gap determine the positions of the peaks
for low frequencies.
The weights of the magnetic peaks are also determined by the 
interaction strength.
For $E_F/\Delta \sim 1$, 
the magnetic peaks form a circle with the radius $\delta$
given by Eq. (\ref{posdel}) which may happen in $BiSCCO$.
However, it is known that the interaction is typically smaller than
the Fermi energy in the superconducting state, so that the condition
for the interaction strength to support the collective mode is not realistic.
Therefore, it suggests that either the charge fluctuations induce
the anomalous interaction or another ingredient is needed 
to explain the incommensurate magnetic peaks found in the experiments.

In the low energy limit, the susceptibilities for the momentum transfer
near ${\bf Q}_0 = (0,0)$ are also interesting.
It is found that
\begin{eqnarray}
{\rm Im} \chi^c_0 (\omega, {\bf Q}) &=&\frac{1}{32\Delta E_F}
 \frac{\omega^2-(\sqrt{2}\Delta q_-)^2} {\sqrt{\omega^2-\omega_q^2}} \ ,
\nonumber\\
{\rm Im} \chi^s_0 (\omega, {\bf Q}) &=&\frac{1}{32\Delta E_F}
\frac{\omega_q^2}{\sqrt{\omega^2-\omega_q^2}} \ ,
\end{eqnarray}
where ${\bf q}={\bf Q}-{\bf Q}_0$.
Note that both the irreducible charge and spin susceptibilities
are singular as $\omega$ approaches $\omega_q$.
However, these singularities in the long wavelength limit are likely 
to be weakened by the vertex corrections.
At realtively high frequencies $\sim 2\Delta$ the charge and spin
collective modes were previously found near ${\bf Q}=(2k_F,0)$ and
$(k_F,k_F)$ respectively.\cite{kee}

Using the above information,
we compute the lowest order contribution to the NMR relaxation rate $1/T_1$
assuming that the scattering near ${\bf Q}_{\pi}$ dominates:
\begin{equation}
\frac {1}{T_1 T} = \lim_{\omega \rightarrow T} 
\int \frac {d^2 Q}{(2 \pi)^2} \frac {{\rm Im} 
\chi^s_0 (\omega, {\bf Q})}{\omega} \simeq 
\frac {T^2}{(\Delta E_F)^2} \ .
\end{equation} 
Thus the relaxation rate $1/T_1$ goes as $T^3$ which is
consistent with the numerical calculations\cite{bulut}
and an experiment for a range of the temperatures\cite{illinoi}.
 
We can also compute the quasiparticle scattering 
rate due to the charge fluctuations.
We consider only the charge fluctuations because these are
 more stronger than those of the spin.
The lowest order self-energy can be obtained from
\begin{equation}
\Sigma = 
i V^2 \int \frac {d^2 Q}{(2 \pi)^2} \int \frac{d\nu}{2\pi}
\frac{(\omega+\nu)I+\xi_{{\bf k}+{\bf Q}}\tau_3+\Delta_{{\bf k}+{\bf Q}}
\tau_1}{(\omega+\nu)^2-E_{{\bf k}+{\bf Q}}^2+i \eta}
\int \frac{dx}{\pi} \frac{{\rm Im} \chi^c_0 (x,{\bf Q})}{\nu-x+i \eta} \ ,
\end{equation}
The imaginary part of the self energy in the $I$ space is 
found at the node, ${\bf k}_0$, as
\begin{eqnarray}
{\rm Im} \Sigma_I(\omega,{\bf k}_0) &=&  V^2  \int_0^{\infty} \frac{dx}{2} 
\int \frac {d^2 Q}{(2 \pi)^2} 
{\rm Im} \chi^c_0 (x,{\bf Q}) \left[\delta(\omega+x+E_{{\bf k}_0+{\bf Q}})
+\delta(\omega-x-E_{{\bf k}_0+{\bf Q}}) \right] \nonumber\\
&\simeq& V^2\frac{\omega^3}{(E_F \Delta)^2} \ .
\end{eqnarray}
It can be shown that the self energies in the $\tau_1$ and $\tau_3$
spaces have larger power in $\omega$, ${\cal O} (\omega^5)$,
compared to $\omega^3$. 
Thus, the leading contribution to the quasiparticle scattering rate, 
$1/\tau(\omega)$, is given by
\begin{equation}
1/\tau (\omega) \simeq V^2 \frac{\omega^3}{(E_F \Delta)^2} \ .
\end{equation}

In summary, we presented the analytic results for the dynamic
charge and spin susceptibilities near ${\bf Q}_{\pi} = (\pi,\pi)$ and
${\bf Q}_0=(0,0)$ in the BCS d-wave superconducting state.
We found that the presence of the d-wave node leads to the
incommensurate charge peaks for a set of wavevectors forming an
ellipse near ${\bf Q}_{\pi}$ and ${\bf Q}_0$. 
We also showed that incommensurate magnetic peaks near ${\bf Q}_{\pi}$
can be obtained
through the random phase approximation if the extremely strong interaction
is assumed.
When the Umklapp process is included, 
higher intensity of the magnetic structure factor would appear at 
$(\pi, \pi (1 \pm \delta))$ and $(\pi (1 \pm \delta), \pi)$
where the incommensurate magnetic peaks were found in the
experiments on $LSCO$ and $YBCO$\cite{cheong,mook,dai2}.
We also compute the NMR relaxation rate going as
$T^3$ and the quasiparticle scattering rate as $\omega^3$
in the superconducting state.

We thank G. Aeppli, S.-W. Cheong, P. Dai, H. Y. Hwang, C. M. Varma,
and R. Walstedt for helpful discussions.
We are especially greatful to Q. Si for explaining his previous 
and recent numerical results.
One of authors (H.Y.K.) thank B. L. Altshuler and C. M. Varma
for motivating  the present problem and D. A. Bonn for explaining 
his experiment.

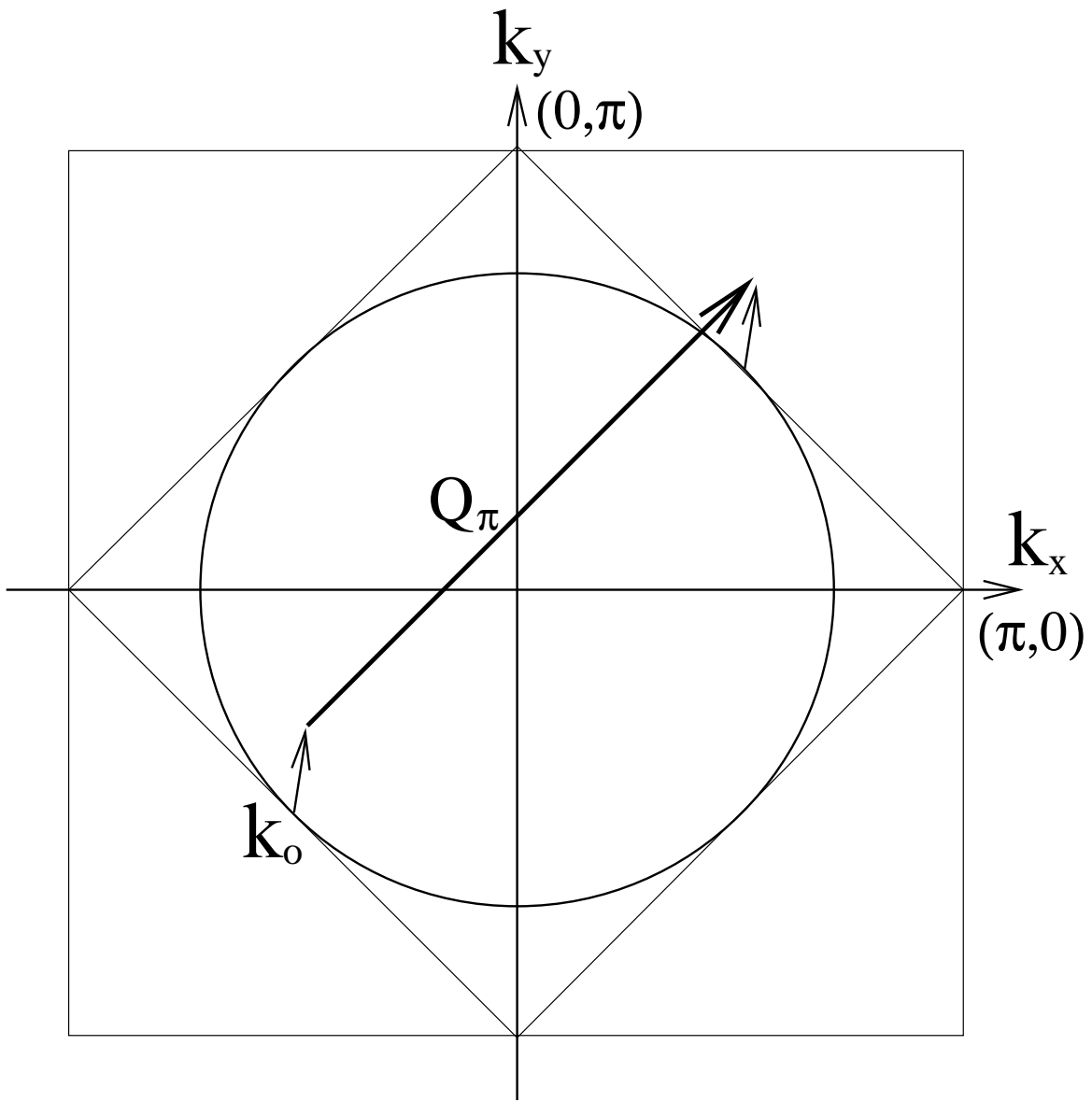
\begin{figure}
\caption{The circular Fermi
surface with a radius $k_F$ illustrating the wave verctor ${\bf Q}_{\pi}$.
The diamond represents the tight binding Fermi surface at the half-filling.
The node positions 
$(\pm k_F/\sqrt{2},\pm k_F/\sqrt{2})$ in the case of the
circular Fermi surface correspond to ${\bf k}_0 = 
(\pm \pi/2, \pm \pi/2)$ in the tight binding model.
As a result, the momentum transfer $(\sqrt{2}k_F,\sqrt{2}k_F)$
corresponds to ${\bf Q}_{\pi} = (\pi,\pi)$. }
\end{figure}

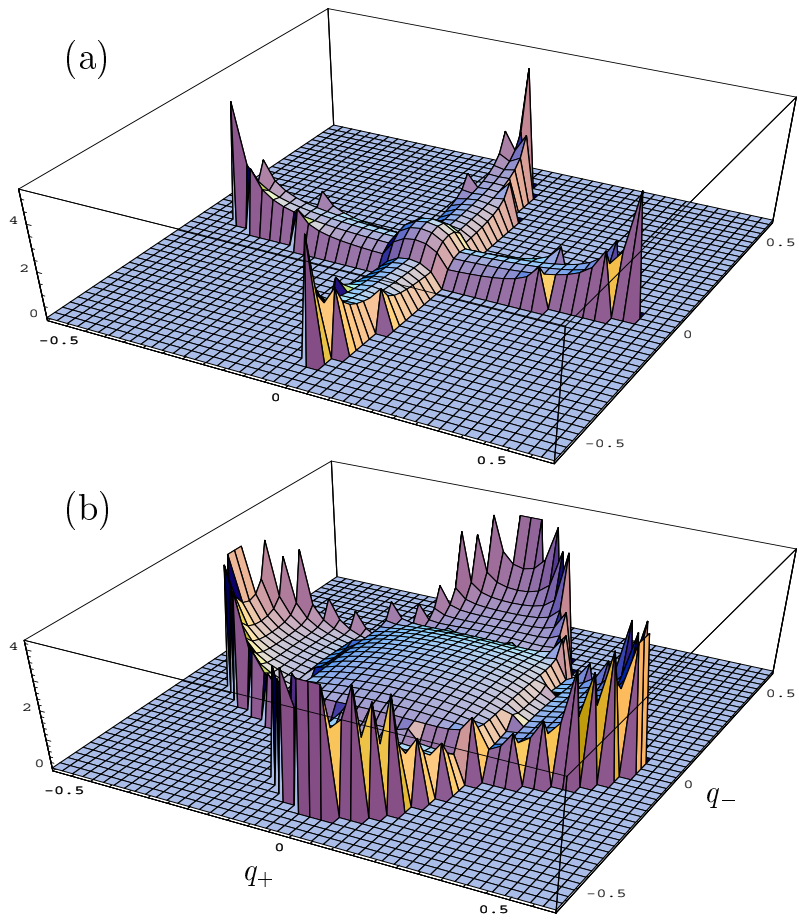
\begin{figure}
\caption{The charge susceptibility for
(a) $E_F/\Delta =7$ and  (b) $2$ with the frequency $0.6 \Delta$.
$q_+$ and $q_-$ are in units of $2\pi/a$.
}
\end{figure}

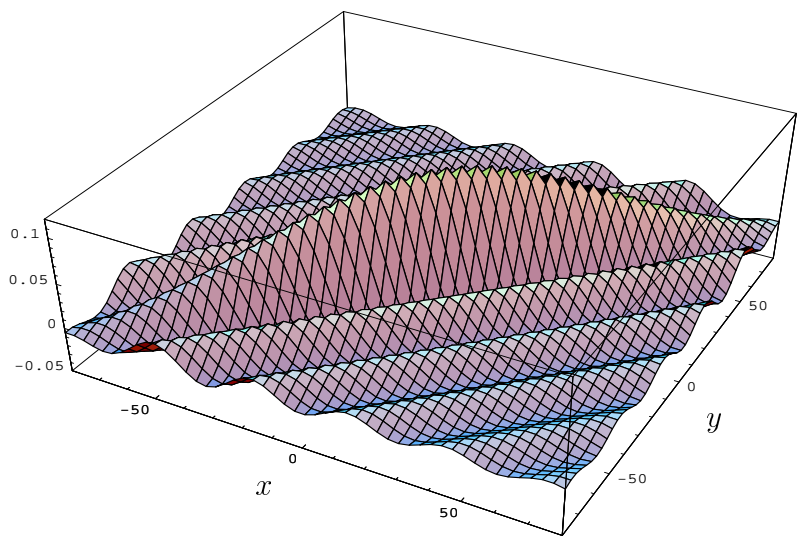
\begin{figure}
\caption{The charge susceptibility in the real space for
$E_F/\Delta =7$ and  the frequency $0.6 \Delta$.
$x$ and $y$ are in units of $a$.
}
\end{figure}


\begin{references}
\bibitem{note1}
As shown in Fig. 1,
$(\pm k_F/\sqrt{2},\pm k_F/\sqrt{2})$ of the circular Fermi surface 
 touches the half-filled Fermi surface near 
$(\pm \pi/2,\pm \pi/2)$.     
Thus, the node positions 
$(\pm k_F/\sqrt{2},\pm k_F/\sqrt{2})$ in the case of the
circular Fermi surface correspond to $(\pm \pi/2, \pm \pi/2)$
in the tight binding model.
As a result, the momentum transfer $(\sqrt{2}k_F,\sqrt{2}k_F)$
corresponds to $(\pi,\pi)$.
\end{references}
\end{document}